\documentclass[prd,aps,superscriptaddress,nofootinbib,showpacs,twocolumn]{revtex4}

\usepackage{graphicx}
\usepackage[usenames,dvipsnames]{color}
\usepackage{amsmath,amssymb}
\usepackage{hyperref}

\newcommand{\vev}[1]{\langle #1 \rangle}
\newcommand{\td}{t_d}

\newcommand{\Gt}{{\widetilde{G}}}
\newcommand{\gt}{{\tilde{g}}}

\newcommand{\GeV}{\,\textrm{GeV}\,}

\newcommand{\paren}[1]{\left( #1 \right)}
\newcommand{\Sigmatot}{\Sigma_{{\textrm{tot}}}}
\def\eprint#1{\href{http://arxiv.org/abs/#1}{#1}} 
\def\gev{{\rm \,Ge\kern-0.125em V}}

\begin{document}

\title{Flat directions and gravitino production in SUSY models}
\author{Raghavan Rangarajan}
\email{raghavan@prl.res.in}
\author{Anjishnu Sarkar}
\email{anjishnu@prl.res.in}
\affiliation{Physical Research Laboratory}
\date{\today}

\begin{abstract}
Flat directions in supersymmetric models can get large vacuum expectation
values in the early Universe which leads to a large mass for gauge bosons
and gauginos.  We point out that this can then result in enhanced
gravitino production because the cross-section for the production of
the $\pm 1/2$ helicity states of the gravitino is proportional to the
square of the gaugino masses.  We consider gravitino production after
inflation in such a scenario and find that the abundance in some cases
can be much larger than the upper bound on the gravtino abundance from
cosmological constraints unless the flat direction field has a very
small vacuum expectation value when it commences oscillating.
\end{abstract}


\maketitle

\section{Flat directions in SUSY models}
\label{sec:flatdir}

Generic supersymmetric (SUSY) models have a large number
of flat directions i.e. directions in the field space
of scalars that have a flat potential \cite{Affleck:1984fy,
Dine:1995uk,Dine:1995kz,Gherghetta:1995dv,Enqvist:2003gh,Dine:2003ax}.
However SUSY breaking and non-renormalizable terms lift these flat
directions and can give rise to a non-zero minimum for the potential of
the field $\phi$ that parametrises a flat direction.  During inflation
the minimum of the potential for $\phi$ can be large.  $\phi$ oscillates
about its potential minimum, which varies with the Hubble parameter
$H$. When $H \sim m_0$, where $m_0$ is of the order of SUSY breaking
scale $\sim 100 \GeV$, the minimum shifts to the origin with $\phi$
displaced from the origin. In certain other cases, the minimum of the
potential during inflation is at the origin yet the flat direction vev
is non-zero due to quantum fluctuations of $\phi$ during inflation.

The flat direction vevs give mass to gauge bosons through terms like
$g^2 A_\mu \tilde{f}^* A^\mu \tilde{f}$ where $\tilde{f}$ represents
a squark or a slepton associated with $\phi$.  The gauge bosons get
a SUSY conserving mass $m_g^2 \sim g^2 \varphi^2$ where $\varphi^2 =
\vev{\phi^* \phi}$.  In this process one or more Standard Model (SM)
gauge symmetries can get broken (at late times $\phi$ decays and gauge
symmetry is restored.)  Gauginos get an equivalent mass, and we take
$m_g = m_\gt = \sqrt{\alpha} \varphi$.

There can be two possible scenarios in the context of flat directions
with large vevs. 

\begin{enumerate}

\item If all the gauge bosons get large masses, then thermalization of
the inflaton decay products is delayed. This is because gauge bosons
mediate the scattering processes relevant for thermalisation and the
scattering cross-section is suppressed due to their large masses.
Lack of thermalization then affects the cosmology of the Universe.
In particular, one needs to recalculate the gravitino production for a
non-thermal Universe, and for a lower final reheat temperature.  This idea
was first proposed in Refs. \cite{Allahverdi:2005fq,Allahverdi:2005mz}.
A more detailed calculation is done in Ref.  \cite{Rangarajan:2012ip}.

\item If not all gauge bosons get mass then thermalization happens as
usual. However, the large vev of a flat direction gives a large mass to
one or more gauginos which can still affect gravitino production. 
\end{enumerate}
    
The gravitino has $\pm 1/2$ and $\pm 3/2$ helicity states and the total
gravitino production cross section is proportional to $\paren{1+m_\gt^2/(3
m_\Gt^2)}$, where the factor ``1" is due to the $\pm 3/2$ helicity states and
the factor $m_{\gt}^2/(3 m_{\Gt}^2)$ is due to the $\pm 1/2$ helicity states
\cite{Bolz:2000fu}.  A large gaugino mass can lead to a large gravitino
abundance in both the above scenarios.  This has been overlooked in the
literature so far. In this article we consider the second scenario in
which only gluons and gluinos get mass and the Universe does thermalize
quickly after inflation.  The first scenario of a non-thermal Universe
after inflation and enhanced gravitino production has been studied in
Ref. \cite{Rangarajan:2012ip}. 

\section{Gravitino production}

Gravitinos are produced by the scattering of the decay products of the 
inflaton. Refs. \cite{Ellis:1984eq,Kawasaki:1994af} provide a list of 
processes for gravitino production. The number density of gravitinos 
generated is then obtained using the Boltzmann equation
\begin{equation}
\frac{dn_{\tilde{G}}}{dt}+3Hn_{\tilde{G}}=
\vev{\Sigmatot|v|} \,n^2 \,,
\label{eq:boltzmann1}
\end{equation}
where $n=(\zeta(3)/\pi^2)T^3$  is the number density of inflaton
decay products ($\zeta(3)=1.20206..$ is the Riemann zeta function
of 3), $\Sigmatot$ is the total scattering cross section for
gravitino production, $v$ is the relative velocity of the incoming
particles, and $\vev{...}$ refers to thermal averaging. We ignore the
gravitino decay term above as the gravitino lifetime is $~10^{7-8}
(100\GeV/m_\Gt)$s \cite{Ellis:1984eq} and is not relevant during the
gravitino production era for gravitinos of mass $10^{2-3}\GeV$ that
we consider.  $\vev{\Sigmatot|v|}$ is given by \cite{Pradler:2006qh}
\begin{align}
\vev{\Sigmatot|v|} &\equiv 
\frac{\eta(T)}{M^2} 
\nonumber \\ &=
\frac{1}{M^2}
\frac{3\pi}{16 \,\zeta(3)} \sum_{i=1}^{3} \left[ 1 
+ \frac{m_{\gt,i}^2(T)}{3 m_{\widetilde{G}}^2}\right] 
c_i g_i^2 
\ln\left( \frac{k_i}{g_i} \right) .
\end{align}
The $i=1,2,3$ in $\eta(T)$ refer to the three gauge groups $U(1)_Y,
SU(2)_L$ and $SU(3)_c$ respectively. $g_i(T)$ are the gauge
coupling constants. The parameters $c_i$ and $k_i$ are constants
associated with the gauge groups and are given by $c_{1,2,3} = 11,
27, 72$ and $k_{1,2,3} = 1.266, 1.312, 1.271$ respectively (see Table
1 of Ref. \cite{Pradler:2006qh}). $M=M_{Pl}/ \sqrt{8\pi}\simeq
2.4\times 10^{18}$ GeV is the reduced Planck mass.  The cross-section
above includes corrections to the expressions obtained earlier in
Refs. \cite{Bolz:2000fu,Kawasaki:2004qu}.  We presume that the inflaton
decays perturbatively and the products thermalise quickly as discussed in
Appendix A of Ref. \cite{Chung:1998rq}.  We shall only consider $i = 3$
because we are studying the enhancement in the gravitino production rate
due to the large gluino mass and we take $\alpha_3 = 6 \times 10^{-2}$
for the relevant energy scales below. Then $\eta(T)$ is given by
\begin{equation}
\eta(T) =
\left[ \frac{3\pi}{16 \,\zeta(3)}
c_3 g_3^2 \ln\paren{\frac{k_3}{g_3}} \right] 
\paren{1+\frac{m_\gt^2}{3 m_\Gt^2}}
\label{eq:eta}
\end{equation}
where $m_\gt$ is now the gluino mass.

To account for the expansion of the Universe one should consider the
abundance of a species $i$ in a comoving volume. This is achieved by
considering the ratio $Y_i=n_i/s$, where $n_i$ is the number density of
particles of the species $i$ in a physical volume and $s$ is the entropy
density given by $s = (2\pi^2/45) \,g_*T^3$, where $g_* = 228.75$ in
the MSSM. Thus Eq. (\ref{eq:boltzmann1}) is rewritten as
\begin{equation}
\dot T\frac{d Y_\Gt}{dT}=\vev{\Sigmatot|v|}
 Y n \,.
\label{eq:boltzmann2}
\end{equation}
In the radiation dominated era, the temperature $T$ is given by
\begin{equation}
T={T_R} \paren{\frac{1}{2H_R(t-t_R)+1}}^{1/2}\,,
\label{eq:T-t-reh}
\end{equation}
where $T_R$ is the reheat temperature and 
\begin{equation}
H_R = \sqrt{ \frac {8\pi^3 g_{*R}}{90} }\frac{T_R^2}{M_{Pl}}\,.
\label{eq:H-reh-value} 
\end{equation}
This implies that $\dot T$ is
\begin{equation}
\dot T =-\frac{H_R}{T_R^2}T^3 = 
-\left( \frac{g_{*R}\pi^2}{90}\right)^{1/2}\frac{T^3}{M}\,.
\end{equation}
Thus
\begin{equation}
\frac{d Y_\Gt}{dT}=-\left( \frac{90}{g_{*R}\pi^2} \right)^{1/2}
\left(\frac{1}{(2\pi^2/45)g_{*}}\right) \left(\frac{\eta}{M}\right)
\left( \frac{\zeta(3)}{\pi^2} \right)^2\,.
\label{eq:YT-boltzmann}
\end{equation}

To obtain the gravitino abundance we integrate the above equation
from $T_R$ to $T_f$ where $f$ corresponds
to the time when the flat direction condensate decays.  Here we
have ignored gravitino production during the period of reheating, as
studied in Refs. \cite{Giudice:1999am,Pradler:2006hh,Kawasaki:2004qu,
Rangarajan:2006xg,Rangarajan:2008zb,Rychkov:2007uq}.  Regarding $t_f$,
the perturbative condensate decay rate is given by $\Gamma_\phi =
m_\phi^3 / \varphi^2$ \cite{Affleck:1984fy,Olive:2006uw} and as argued
in Ref. \cite{Olive:2006uw} the condensate decay products do not dominate
the Universe for $\varphi < 10^{-2} M_{Pl}$ for gravitational decay of the
inflaton.  (We discuss alternate mechanisms for condensate decay below.)
For gravitino production after $t_f$ there will be no enhancement due
to a large gluino mass and the abundance generated after $t_f$ will be
proportional to $T_f$ as similar to the standard scenario.

We take the inflaton mass $m_\psi$ to be $10^{13} \GeV$ 
and the inflaton decay rate
$\Gamma_d \sim m_\psi^3/M_{Pl}^2 \sim 10 \GeV$.  Below we are primarily
concerned with the evolution of $\phi$ after $t_0\sim m_0^{-1}$.  In the
both the cases mentioned at the beginning of Sec. (\ref{sec:flatdir})
$\phi$ effectively has a quadratic potential after $t_0$ with a minimum at
the origin with a positive curvature of $m_0^2$ (ignoring thermal effects
for now).  The gluino mass is given by $m_\gt^2 = \alpha \varphi^2$,
and $\varphi$ is $\varphi_0$ at $t_0$ and then falls as $1/a^{3/2}$
once the condensate starts oscillating at $t_0$.  Then for  $t > \td$,
where $t_d = \Gamma_d^{-1}$ is the inflaton decay time, the gluino mass is
\begin{align}
m_\gt^2 &= \alpha \varphi_0^2 \paren{\frac{a_0}{a}}^3 
\nonumber \\ &=
\alpha \varphi_0^2 \paren{\frac{a_0}{a_d}}^3 \paren{\frac{a_d}{a}}^3
\nonumber \\ &= 
\alpha \varphi_0^2 \paren{\frac{\Gamma_d}{m_0}}^2 \paren{\frac{T}{T_R}}^3 \,,
\label{eq:gluinomass_m0gGd}
\end{align}
where we have used $a \sim t^{2/3}$ for $t_0 < t < t_d$ for an inflaton
oscillating in a quadratic potential during reheating and $a \sim 1/T$
for $t > t_d$.

\begin{widetext}
Integrating the Boltzmann equation in Eq.  (\ref{eq:YT-boltzmann}) from
$T_R$ to $T_f$ gives the gravitino abundance generated between $t_d$
and $t_f$ as
\begin{align}
Y_\Gt &=
\left[ 
\paren{\frac{90}{g_{*R} \pi^2}}^{1/2}
\paren{\frac{45}{2 \pi^2 g_{*R}}} 
\paren{\frac{\sqrt{8 \pi} }{M_{Pl}}}
\paren{\frac{\zeta (3)}{\pi^2}}^2 \right]
\left[ \frac{3\pi c_3 }{16 \,\zeta(3)} 
\right]
\nonumber \\ &
\quad {}\times
\,g_3^2(T_R) \ln\paren{\frac{k_3}{g_3(T_R)}} 
\paren{ (T_R - T_f) + \frac{1}{4} \frac{m_\gt^2(T_R)}{3 m_\Gt^2} T_R
\paren{1-\frac{T_f^4}{T_R^4}}} \,.
\label{eq:Ym0gGd}
\end{align}
We have ignored the variation of $g_*$ and $g_i$ with temperature and
have used the value at $T_R$, since for $T_R \gg T_f$ most gravitino
production occurs close to $T_R$.  The term proportional to $T_R-T_f$ is
associated with the production of $\pm 3/2$ helicity gravitinos and
the other term is associated with the production of the $\pm 1/2$
helicity states.
\end{widetext}

% For $m_0 > \Gamma_d$
The reheat temperature is given by $T_R = 0.55 \,g_*^{-1/4} (M_{Pl}
\Gamma_d)^{1/2} = 2\times 10^{9} \GeV$ \cite{Kolb:1988aj}.
The condensate decays when its decay rate $\Gamma=m_0^3/\varphi^2$
equals $H$, i.e., when $t_f = \varphi_f^2/m_0^3$,
where
\begin{align}
\varphi_f 
&
    = \varphi_0 \paren{\frac{a_0}{a_f}}^{3/2}
\nonumber\\ &
    = \varphi_0 \paren{\frac{a_0}{a_d}}^{3/2}
    \paren{\frac{a_d}{a_f}}^{3/2}
\nonumber \\ &
    = \varphi_0 \paren{\frac{\Gamma_d}{m_0}}
    \paren{\frac{m_0^3}{\Gamma_d \,\varphi_f^2}}^{3/4} \,.
\end{align}
Here we have used $a \sim t^{2/3}$ for $t_0 < t < t_d$ and $a\sim t^{1/2}$ 
for $t > t_d$. This implies
\begin{align}
\varphi_f &
    = \varphi_0^{2/5} \, m_0^{1/2} \, \Gamma_d^{1/10} \,.
\label{eq:phif_m0gGd}
\end{align}
The temperature at $t_f$, using $a \sim t^{1/2}$ for $t>t_d$, is
\begin{align}
    T_f &= T_R \paren{\frac{a_d}{a_f}}
\nonumber \\
    &= T_R \paren{\frac{m_0^3}{\Gamma_d \varphi_f^2}}^{1/2}
\nonumber \\
    &= T_R \paren{\frac{m_0}{\varphi_0^{2/5} \, \Gamma_d^{3/5}}}
\nonumber \\
    &= 0.55 \, g_*^{-1/4} \frac{ m_0 \, M_{Pl}^{1/2}}
        { \varphi_0^{2/5} \,\Gamma_d^{1/10} } \,.
\label{eq:Tf}
\end{align}
For the condensate to decay after inflaton decay, or $T_f < T_R$, we
need $\varphi_0>30 \,m_0$ or $\varphi_f>3 \,m_0$, for $\Gamma_d=10\GeV$.

\section{Results and conclusion}

A large gravitino abundance can be in conflict with cosmological
observations. If the gravitino is stable its energy density can overclose
the Universe. If the gravitino is unstable its decay products can either
overclose the Universe or dissociate the light nuclei generated during
primordial nucleosynthesis.  We use an upper bound of $Y_\Gt < 10^{-12}$ for
consistency with cosmological observations. For the abundance obtained
in Eq.  (\ref{eq:Ym0gGd}) to be consistent with observations then requires
\begin{equation}
\frac{\varphi_0}{m_\Gt} < 300 \,.
\label{eq:varphi0ubnd}
\end{equation}
For $m_\Gt = 100 \GeV$, the constraint is 
\begin{equation}
\varphi_0 < 3\times10^4 \GeV \,.
\label{eq:varphi0bnd}
\end{equation}
Note that if we fix  $m_\Gt = 100 \GeV$ and allow the condensate mass,
so far taken to be $m_0$, to vary one gets the same constraint
as in Eq. \ref{eq:varphi0ubnd} with $m_\Gt$ replaced by the
condensate mass $m_\phi$ (since in Eq. (\ref{eq:Ym0gGd}) 
$m_\gt^2/m_\Gt^2\sim\varphi_0^2/(m_0^2 m_\Gt^2)$).
This is relevant if one includes thermal corrections to the condensate
potential.

We obtain this bound assuming $T_f \ll T_R$ in Eq. (\ref{eq:Ym0gGd}). 
From Eq. (\ref{eq:Tf}) $\varphi_0=3\times10^4 \GeV$ implies 
$T_f = 6 \times 10^8 \GeV$. For this value of $T_f$ and $T_R = 2\times 10^9 \GeV$, 
our assumption only slightly affects the contribution of the first term 
in Eq. (\ref{eq:Ym0gGd}), and does not alter the contribution of the 
second term. Since the contribution of the second term is larger than 
that of the first term our assumption is therefore justified. Smaller 
values of $\varphi_0$ will correspond to larger $T_f$ and therefore greater 
compatibility of $Y_\Gt$ with cosmological constraints.

We now consider plausible values of $\varphi_0$.
During inflation the non-zero vacuum energy breaks SUSY and can give
large positive masses to the flat direction of order $H_I$, where
$H_I$ is the Hubble parameter during inflation \cite{Dine:1995uk}.
This may be true in supergravity models with minimal or non-minimal
Kahler potentials \cite{Gaillard:1995az}. The flat direction potential
has a minimum at the origin and the field is driven to this minimum.
Quantum fluctuations during inflation then give a vev of order $H_I$
\cite{Dine:1995uk}.  Assuming that the field does not vary much till
$t_0$, then $\varphi_0\sim H_I$ and the upper bound on $\varphi_0$
implies that the inflationary scale $V_I^{1/4}<4\times10^{11}\GeV$.

Alternatively, 
in no scale supergravity or more generally in any supergravity theory with
a Heisenberg symmetry of the kinetic function in the Kahler potential,
one gets contributions during inflation to the flat direction potential
from supersymmetry breaking (due to $H_I$) only at the one loop level
and these are negative at the origin for flat directions which do not
involve a stop and if the inflaton is not the standard string dilaton
\cite{Gaillard:1995az}.  This correction to the potential, along with
non-renomalizable terms, leads to a shifted minimum of the potential
and subsequently a large vev of order $M_{Pl}$ \cite{Gaillard:1995az}
or $10^{12-14} \GeV$ on including GUT interactions \cite{Campbell:1998yi}
as the field rolls to the minimum of its potential 
%\cite{Gaillard:1995az}
and oscillates about it.  (Negative corrections at the origin were also
discussed in Ref. \cite{Dine:1995uk}.)

We consider the shifted 
minimum of the potential for $\phi$ to be $M(H/M)^{1/n+1}$
where non-renormalizable terms in the potential are of the form
$\phi^{2n+4}/M^{2n}, n \ge 1$ \cite{Campbell:1998yi,Dine:2003ax}. $\phi$
oscillates about this time dependent minimum which decreases as $H$
decreases.  When $H \sim m_0$, the potential minimum goes to zero and
the field oscillates about the origin in a quadratic potential with
curvature $m_0^2$.  Then \cite{Dine:2003ax}
\begin{align}
\varphi_0 &\sim
M ( H(t_0) / M )^{1/(n+1)} \,, 
\label{eq:varphi0M}
\end{align}
where $H(t_0) = 100 \GeV$. 
%The least $\varphi_0$ is obtained for 
For $n=1$, $\varphi_0 < 3 \times 10^4 \GeV$ is satisfied for 
\begin{equation}
M < 10^7 \GeV.
\end{equation}
For larger values of $n$, the upper bound on $M$ reduces further
and one requires $M<10^{4-5}\GeV$ for $n>1$.  However we are
considering scales upto at least $T_R \sim 10^9 \gev$, and
setting $M=10^9\gev$ gives a value
of $\varphi_0$ in conflict with the upper bound on $\varphi_0$.  If we
choose $\varphi_0 \sim 10^{12}\GeV$ as mentioned above
then we get the gravitino abundance
$Y_\Gt$ to be $8\times10^{2}$ which is orders of magnitude higher than
the cosmological bound of $10^{-12}$.  The abundance will increase as
$\varphi_0^2$ for larger values of $\varphi_0$.

Thermal corrections to the condensate potential can increase the mass of
the condensate field \cite{Allahverdi:2000zd,Anisimov:2000wx,Anisimov:2001dp}.
Ref. \cite{Anisimov:2001dp} obtains masses of order $10^{10} \GeV$,
$10^6 \GeV$ and $10^5 \GeV$ for non-renormalizable terms with $n = 1,2,3$
and $M = 10^{18} \GeV$ . We find that the condition $\varphi_0 <
300\, m_\phi$ discussed above is not satisfied for $\varphi_0$ obtained
from Eq. \eqref{eq:varphi0M} with $H(t_0)$ replaced by $m_\phi$.

Our results imply that either one has a scenario of supergravity with
a positive contribution to the $\phi$ potential from $H_I$ that gives
$H_I<3\times10^4\GeV$ and $V_I^{1/4}<4\times10^{11}\GeV$,
or the flat directions must decay quickly in a way so as to not
affect the cosmology of the Universe. 

With regards to the quick decay of the flat directions,
the longevity of flat directions has been debated in
Refs. \cite{Allahverdi:1999je,Postma:2003gc,Olive:2006uw,
Allahverdi:2006xh,Basboll:2007vt,Basboll:2008gc,Allahverdi:2008pf,
Gumrukcuoglu:2008fk,Gumrukcuoglu:2009fj,Allahverdi:2010xz}.  However it
has been argued in Refs. \cite{Allahverdi:2008pf,Allahverdi:2010xz} that
even if non-perturbative rapid decay via parametric resonance occurs for
scenarios with multiple flat directions it leads to a redistribution of
energy of the condensate amongst the fields in the D flat superspace
and hence to practically the same cosmological consequences, including
at least as large gauge boson and gaugino masses as in the scenario
with only perturbative decay.

Scattering of particles of the thermal bath off the flat direction
condensate can lead to the decay of the condensate \cite{Dine:1995kz,
Allahverdi:2000zd,Anisimov:2000wx}, though thermal
effects are less important for larger values of $n$ . For example, for $n=3$ 
the condensate
decays much after the decay of the inflaton \cite{Anisimov:2000wx}. Decay
via fragmentation into solitonic states called Q-balls
\cite{Kusenko:1997si,Enqvist:1997si,Kasuya:1999wu,Enqvist:2000gq,
Kasuya:2000sc,Kasuya:2000wx,Kasuya:2001hg, Enqvist:2000cq,Multamaki:2002hv}
or Q-axitons \cite{Enqvist:1999mv} due to inhomogeneities in the
condensate may also be relevant. However the time scales for the
formation of Q-balls and Q-axitons can be larger than $t_f$. $t_f$
is less than $60 \,m_0^{-1}$, for $\varphi_f < 800 \GeV$ obtained from
Eq. (\ref{eq:phif_m0gGd}).

In conclusion, the presence of flat directions in SUSY models and
associated large vevs seems to have very important consequences for
gravitino production in the early Universe.  These large vevs give a
large mass to gauginos which enhances gravitino production and can violate
cosmological constraints on the gravitino abundance.  This scenario can be
avoided if the flat direction has a vev smaller than $ 3 \times 10^4 \GeV$
at $t\sim m_0^{-1}$, or if the flat direction decays early.

% \bibliographystyle{apsrev}
% \bibliography{cosmology,genesis,susy,qballs}

\end{document}